# Manipulation of Kondo Effect via Two-Dimensional Molecular Self-Assembly

*Violeta Iancu, Aparna Deshpande, Saw-Wai Hla\**

*Nanoscale & Quantum Phenomena Institute, Department of Physics and Astronomy, Ohio University, Athens, Ohio, 45701, USA*

**We report manipulation of a Kondo resonance originated from the spin-electron interactions between a two-dimensional molecular assembly of TBrPP-Co molecules and a Cu(111) surface at 4.6 K using a low temperature scanning tunneling microscope. By manipulating nearest-neighbor molecules with a scanning tunneling microscope tip we are able to tune the spin-electron coupling of the center molecule inside a small hexagonal molecular assembly in a controlled step-by-step manner. The Kondo temperature increases from 105 to 170 K with a decreasing the number of nearest neighbor molecules from six to zero. This Kondo temperature variation is originated from the scattering of surface electrons by the molecules located at the edges of the molecular layer, which reduces spin-electron coupling strength for the molecules inside the layer. Investigations on different molecular arrangements indicate that the observed Kondo resonance is independent on the molecular lattice.**

[*]Corresponding author, Email: hla@ohio.edu, web: www.phy.ohiou.edu/~hla

A control over spin-electron interactions is vital for development of spintronic devices and for quantum computation [1-11]. When a magnetic impurity is surrounded by free electrons, a realignment of the electron spins occurs below a critical temperature due to spin-electron interactions --a phenomenon known as the "Kondo" effect [12]. The Kondo effect has been observed in a wide range of systems including single atoms/molecules [1-9], quantum-dots [10], and carbon nanotubes [11], however two-dimensional molecular Kondo systems have yet to be explored. Molecules with magnetic properties recently have great appeal as they offer an ideal platform to advance the fundamental understanding of spin related mechanisms [5-9], and can act as templates for molecular spintronic device fabrication due to their propensity for spontaneous self assembly. The molecule of interest in our study, TBrPP-Co [5, 10, 15, 20 –Tetrakis -(4-bromophenyl)-porphyrin-Co], is composed of a porphyrin unit with a cobalt (Co) atom caged at its center and four bromo-phenyl groups at the end parts [9].

For this study, a custom-built low-temperature scanning tunnelling microscope system [13] operated in ultrahigh vacuum, typically at a pressure below 4 x $10^{-11}$ Torr, was used. A Cu(111) sample was cleaned by repeated cycles of Ne ion sputtering and annealing to 800 K. A sub-monolayer coverage of molecules was deposited using a home-built Knudson cell on an atomically clean Cu(111) surface held at ~ 120 K. The sample temperature was further lowered to 4.6 K for the measurements. For a comparative study, TBrPP-Cu molecules, where the cobalt atom is replaced by a copper atom, were separately deposited on Cu(111).

Both TBrPP-Co and TBrPP-Cu molecules self assemble to form ribbon-like monolayer islands with a preferential growth direction along a 7º deviation from the [110] surface directions (Fig. 1a, 1b). Both molecule types can adsorb on Cu(111) with two molecular conformations; planar and saddle [9]. Within the ribbon assembly, the molecules form a hexagonal arrangement (Fig. 1c) and have a planar conformation [9] with their porphyrin planes lying parallel to the surface. The molecules having saddle conformations do not self-assemble at this sample preparation condition and are sporadically located on the bare Cu(111) surface. Thus we focus our attention on the two dimensional assembly of planar TBrPP-Co only in this study.

Scanning tunnelling microscope (STM) is an ideal instrument to investigate the spin-electron interaction at an atomic limit on surfaces by measuring the energy dependence of the local density of states around the Fermi level [1-6, 9]. The Kondo temperature can be determined from the shape and width of the resonance

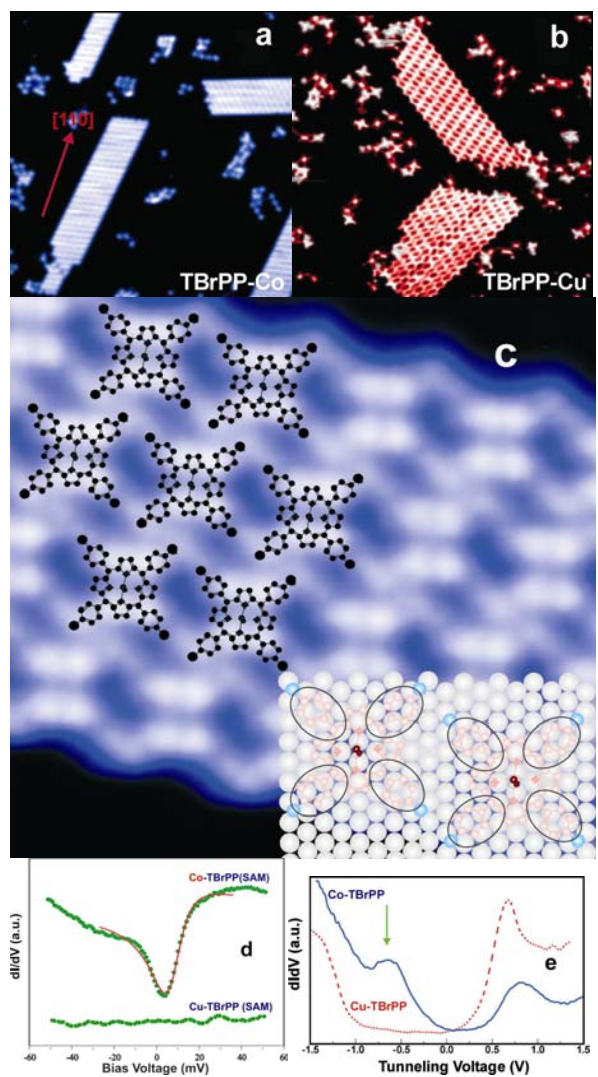

**FIG 1.** Self-assembled molecular ribbons. STM images of TBrPP-Co **(a)**, and TBrPP-Cu **(b)**, on Cu(111)showing ribbon-like self-assembled molecular assemblies. At higher resolution, a hexagonal close-packing of the molecules can be observed **(c)**. The distance between the centers of two nearest neighbor molecules is 1.7 nm (Imaging parameters: 1V, 0.25nA). **(d)** A differential conductance spectrum of a TBrPP-Co molecule inside the ribbon shows a dip (Kondo resonance) while that of TBrPP-Cu does not show any resonance around the surface Fermi level (0V). The solid line represents the Fano line-shape fit to the data. The fit parameters are q = 0.38 ± 0.01 and $\varepsilon_K$ = 7.11 ± 0.12 mV **(e)** The HOMO located at 0.7eV (indicated with an arrow) is originated from the Co $d_z^2$ orbital (blue curve). This orbital is absent in the OMTS of TBrPP-Cu ribbon (red curve). The peaks located at the positive bias are the lowest unoccupied orbitals of TBrPP-Co and TBrPP-Cu molecules.

[1,2,9] located around 0 V (Fermi level) of the differential conductance (dI/dV) tunneling spectra. To investigate the Kondo effect generated from the two-

dimensional molecular layer, the dI/dV tunneling spectra are acquired by positioning the STM-tip above the center of a molecule inside the ribbons formed by TBrPP-Co and TBrPP-Cu, respectively. The dI/dV spectra of TBrPP-Co reveal a resonance located around the substrate Fermi level, while that of TBrPP-Cu show no apparent features (Fig. 1d). Since the TBrPP-Co includes a spin-active cobalt atom, we attribute the observed resonance as caused by the Kondo effect. In agreement with the observed Kondo effect, the orbital mediated tunnelling (dI/dV vs. V) spectra [9] taken at the center of a TBrPP-Co molecule from a TBrPP-Co ribbon show an occupied molecular orbital located at 0.7 eV below the Fermi level. The origin of this orbital is assigned to the ionisation of $d_z^2$ orbital of the Co atom [9, 14], which is absent in case of TBrPP-Cu (Fig. 1e). The Kondo effect in TBrPP-Co molecular assembly is repeatedly measured with different tips on a large number of molecules for consistency. The experimental data is fitted by using the following formula [15];

$$\delta_{\rho_R}(\omega) = \frac{[\mathrm{Im}\, G^{(0)}{}_{R,\sigma}(\omega - i\delta)]^2}{\pi \rho_0} \times \left\{ \frac{2 q_R \varepsilon + q_R^2 - 1}{\varepsilon^2 + 1} + C_R \right\};$$

$$\varepsilon = (\omega - \varepsilon_K)/T_K,$$

where q, $T_K$, and $\varepsilon_K$ represent the line-shape parameter, the Kondo temperature, and the energy shift of the Kondo resonance, respectively. The average measured Kondo temperature is 105 ± 10 K, and the error here represents a slight deviation of $T_K$ measured with the different tips.

To manipulate the observed Kondo resonance, we use an innovative approach: First a small molecular assembly, a 'hexagon', is created by removing molecules from a TBrPP-Co ribbon with the STM-tip (Fig. 2a, and 2b) (see the STM movie for the Hexagon creation [20]). Next, this 'hexagon' is disassembled by moving one molecule-at-a-time using STM manipulation [13]. After each molecule removal, the corresponding Kondo resonance is determined above the center molecule of the 'hexagon'.

Initially, the center molecule is completely surrounded by six molecules and the Kondo temperature is measured as 105 K (Fig. 3). This value is the same as the ones measured inside full ribbon structures indicating that the second or third nearest neighbors in the molecular assembly do not play a key role for the observed Kondo resonance. Next, one of the six surrounding molecules from the 'hexagon' is removed and the Kondo signature is determined above the same center molecule again (Fig. 3a). This time the dI/dV signal shows an increase in the resonance width (Fig. 3b, and 3c). The width of the Kondo resonance is a measure of the Kondo temperature and therefore, a measure of the spin-electron coupling strength. The increased Kondo temperature means that the spin-electron coupling is now enhanced. Continue reduction in the number of nearest neighbor ('nn') molecules increases in the Kondo temperature of the center molecule further. Finally, for the isolated molecule (at the '0 nn'), the Kondo temperature reaches to 170 K. Thus, using this manipulation scheme, we are able to tune the Kondo temperature between 105 and 170 K, which amounts to ~38% temperature variation.

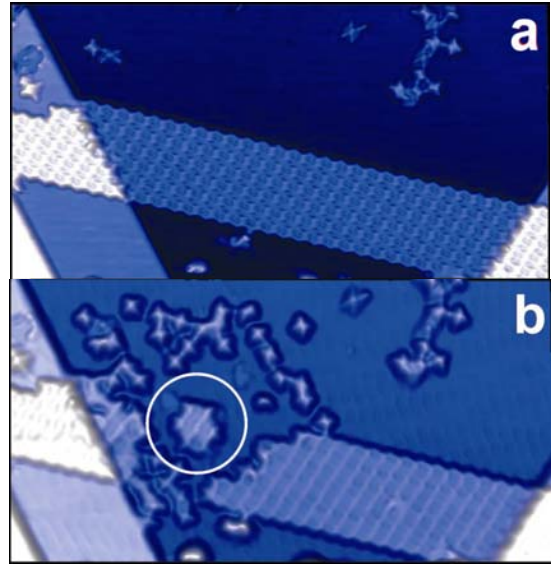

**FIG 2.** Building molecular 'Hexagon'. The molecules from the ribbon **(a)** are removed by using STM tip to create a hexagonal unit cell (circled) **(b)** [Imaging parameters: 28 nm x 50 nm, 1 V, 0.2 nA]. (See the STM movie for the Hexagon creation [20])

There are several possible mechanisms that might contribute to the observed Kondo temperature variation. The Kondo temperature '$T_K$' is related to the density of the conducting electrons 'ρ' and the exchange coupling 'J' at the magnetic impurity [16,17] as: $T_K \propto \exp[-(1/\rho J)]$. Thus, variation of 'ρ' or 'J' or both would change the Kondo temperature. The possible factors that can significantly change 'ρ' and 'J' in this case are; 1) formation of spin network, 2) hybridization of molecules within the molecular layer, 3) distortion of molecules due to the packing, and 4) changes of surface electron density due to molecular assembly.

When TBrPP-Co is deposited on Cu(111) at ~90 K, two novel self-assemblies can be formed in addition to the ribbon structure. At the first structure, TBrPP-Co assembles in a parallel arrangement with the unit cell distances of 1.7 and 1.5 nm (Fig. 4a). At the second structure, the molecules are only partially ordered and the basic molecular unit is formed with a triangular arrangement of three molecules rotated by 120º with molecule-at-a-time with the STM-tip (left) from the 'hexagon' and corresponding models (middle). The calculated electron standing wave patterns reveal a gradual exposing of the center molecule (indicated with a red circle) to the surface state electrons. Here white and black colors in the calculated images represent higher and lower electron densities, respectively. The black region under the molecular clusters indicates reduction of surface electronic charge density. **(b)** The dI/dV spectra are measured at each step by positioning the tip above the center molecule (indicated with red dot). The spectra are vertically and horizontally displaced for clarity. Horizontal displacements of 3 to 10 meV are taken for the spectra representing nn6, nn5, nn4, nn2 and nn1. **(c)** The plot of Kondo temperature as a function of the number of nearest neighbors.

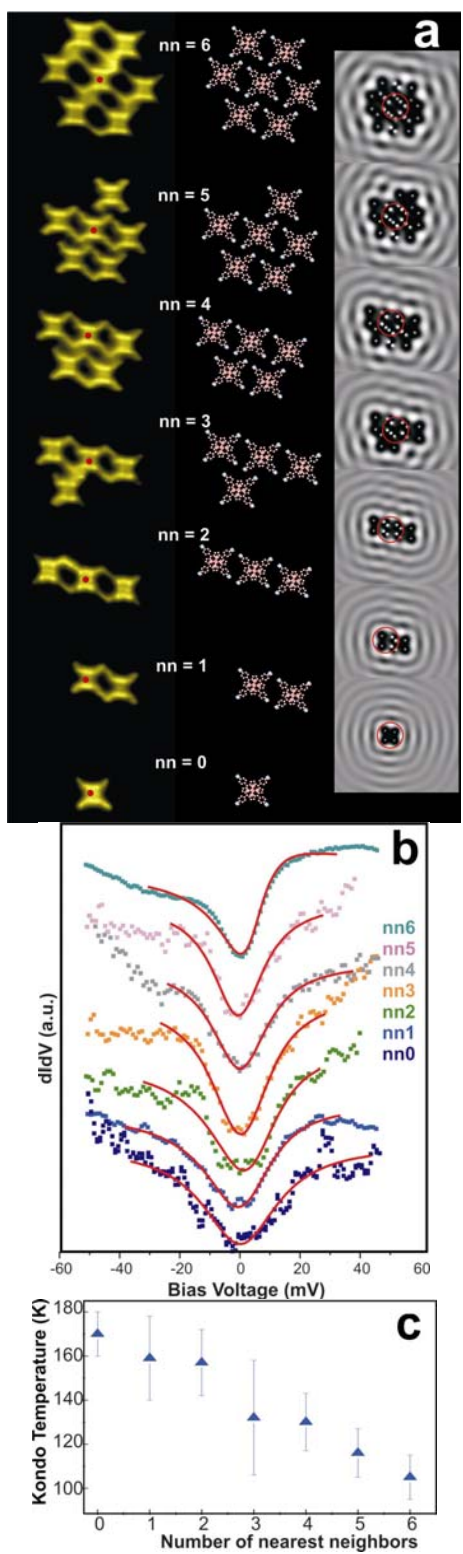

**FIG 3.** Kondo temperature tuning. **(a)** A sequence of STM images of different 'nn' molecules created by removing one-

respect to each other. The distance between the molecules here is 1.7 nm (Fig. 4b). Based on the differences in lattice arrangements, these two molecular assemblies can be used as viable experimental platforms to test the effect of spin network on the observed molecular Kondo effect. The Kondo signatures of these two structures are determined by using the same procedure as above and found to be 108 ± 12 K for the parallel arrangement, and 111 ± 21 K for the triangular arrangement, respectively. Within the error bar, these values are identical to that of the ribbon. Since there are no apparent changes in Kondo temperatures of the molecular assemblies with different lattice arrangements and distances, the contribution of the two-dimensional spin network as well as the possible hybridization inside the molecular layer to the Kondo effect could be excluded. If the molecules are distorted due to the packing arrangement, their orbital locations should alter. The main contributing orbital, $d_z^2$, to this process remains the same (0.7 eV below the Fermi level) for the single molecule [9] and molecular layer (Fig. 1e) indicating that the distortion effect can also be ruled out.

For single Co atoms adsorbed on Cu(111), Knorr et al. [16] have shown that the Kondo effect is mainly contributed from the bulk electronic states even though the surface states are still accounted for the spin-electron coupling [3]. Following this work, Barral et al [18] showed that both surface and bulk electronic states are involved in the exchange spin-electron coupling of this system, however, in agreement with the Knorr et al report, bulk states contribution is higher than that of the surface states. TBrPP-Co is a relatively large molecule composed of 77 atoms that form four bromophenyl groups in addition to a center porphyrin unit and thus, it covers a relatively large surface area (~ 1.5 nm x 1.6 nm) [9]. Therefore, it is necessary to consider both, the coupling of Co atom caged at the molecule's center as well as that of the molecular ligands to the surface. Due to its large size as compared to a single Co atom, the spin-electron coupling involving both surface and bulk electronic states cannot be ruled out in this TBrPP-Co-Cu(111) system.

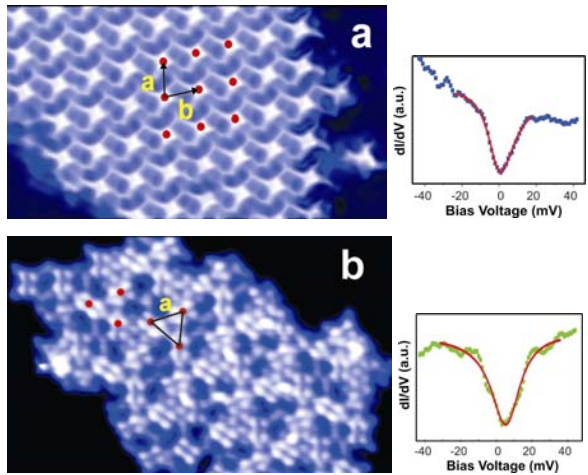

**FIG 4.** Different TBrPP-Co self-assemblies. (a), TBrPP-Co assemble in a parallel arrangement with lattice spacing a = 1.7, b = 1.5 nm. The kondo temperature of 108 ± 12 K is determined from the dI/dV tunnelling spectroscopy (right, q = - 0.16 ± 0.01 and $\varepsilon_K$ = -0.4 ± 0.2 meV). (b), In this TBrPP-Co cluster, the molecules assemble in a partial order of triangular arrangement. The triangular lattice has a = 1.7 nm. The Kondo temperature of 111 ± 21 K is measured (q = 0.06 ± 0.01 and $\varepsilon_K$ = 5.8 ± 0.4 meV) from the dI/dV data (right).
[STM imaging parameters: 1V, 0.4 nA. STM image size: 15 nm x 9 nm (a) and 17 nm x 10 nm (b)]

In their theoretical work, Barral et al. [16] have proposed that the Shockley surface state in Cu(111) is quenched if a two dimensional network of adatoms are formed with 10 or less atomic distances, i.e. < 2.54 nm, between the adatoms. The quenching of surface state decreases the electron density 'ρ' and thus it can reduce the Kondo temperature. To check the electronic density under the cluster, we have calculated the surface electron standing wave patterns for the molecular clusters having different number of molecules on Cu(111) and the calculated images are displayed in Fig. 3a [19]. In these images, no electron standing wave pattern can be observed at the area under the molecular cluster. Since the standing waves are generated by the scattering of surface state electrons, depletion of standing waves indicates a reduced density of surface state electrons under the hexagon. It appears that the scattering of surface electrons is mainly caused by the molecules located at the edges of the cluster. Thus, the molecule located inside the cluster, where the Kondo temperature is measured, will have less interaction with the surface state electrons as compared to the ones at the edges. At the full hexagon, i.e. '6nn' configuration, the center molecule (indicated with a red circle in the calculated images in Fig. 3a) is completely surrounded by six neighbors. By removing the nearest neighbor molecules one-at-a-time, from '6' to '0', the center molecule is increasingly exposed to the surface state electrons. This should increase 'ρ' and thus, the associated Kondo temperature should rise.

To further verify the mechanism, we measure the Kondo resonance over the molecules located at the sides, edges and corners of the three molecular assemblies. For instance, the molecules located at the sides of the clusters can have either '3nn' or '4nn' while that at the corner can have '2nn' (Fig1c, Fig. 4a, and 4b). Here, the measured Kondo signatures agree well with the values obtained from the manipulation experiment having the same nearest neighbors.

In summary, we systematically show that by varying the nearest neighbor coordination number of molecules inside a two dimensional molecular self-assembly, the spin-electron coupling strength can be manipulated and consequently, the Kondo temperature can be tuned in a controlled manner between 105 K and 170 K. This manipulation of two-dimensional molecular Kondo effect merges spintronic research area to a rapidly evolving and the most promising field of nanoscience, molecular self-assembly, and thus it opens novel routes for 'molecular spintronic' applications.

We thank L. Dias da Silva, and S.E. Ulloa for thoughtful discussions and K.-F. Braun for providing the standing wave program. We would like to acknowledge the Ohio University Bionano Technology Initiative and funding provided by the United States Department of Energy, BES grant number DE-FG02-02ER46012.